# The $Ba_2LnFeNb_4O_{15}$ "Tetragonal Tungsten Bronze": towards RT composite multiferroics

M. Josse[a,*], O. Bidault[b], F. Roulland[a], E. Castel[a], A. Simon[a], D. Michau[a], R. Von der Mühll[a], O. Nguyen[a] and M. Maglione[a]

[a] ICMCB-CNRS, Université Bordeaux 1, 33608 Pessac, France
[b] I.C.B., CNRS - Université de Bourgogne, 21078 Dijon, France

[*] To whom correspondence should be addressed: josse@icmcb-bordeaux.cnrs.fr

# Abstract

Several Niobium oxides of formula $Ba_2LnFeNb_4O_{15}$ (Ln = La, Pr, Nd, Sm, Eu, Gd) with the "Tetragonal Tungsten Bronze" (TTB) structure have been synthesised by conventional solid-state methods. The Neodymium, Samarium and Europium compounds are ferroelectric with Curie temperature ranging from 320 to 440K. The Praseodymium and Gadolinium compounds behave as relaxors below 170 and 300 K respectively. The Praseodymium, Neodymium, Samarium, Europium and Gadolinium compounds exhibit magnetic hysteresis loops at room temperature originating from traces of a barium ferrite secondary phase. The presence of both ferroelectric and magnetic hysteresis loops at room temperature allows considering these materials as composites multiferroic. Based on crystal-chemical analysis we propose some relationships between the introduction of $Ln^{3+}$ ions in the TTB framework and the chemical, structural and physical properties of these materials.

# Introduction

Much attention has been paid in the recent years to multiferroic materials, and particularly to those presenting both ferromagnetism and ferroelectricity. Magnetoelectric multiferroics, which will be referred to as multiferroics in the following, could be used, for example, for the design of new generation of Random Access Memories (RAMs) in which both the electric and magnetic polarisation are used for data storage. Aside from the processing of multiferroic thin films and composites [1], the search for bulk multiferroic is a great challenge.

However up to now the candidate multiferroic materials for room temperature applications are very few. A large majority of the "multiferroic" publications of the recent years deal with perovskite-related materials [2,3]. From this abundant literature, two limitations are widely accepted. First, the electronic structure of the A an B cations in $ABO_3$ multiferroics is hardly compatible with a magnetic ordering at elevated temperatures [4]. Second, a coupling between the ferroelectric and magnetic order may be obtained when the material exhibit a non standard, e.g. spiral, magnetic structure [5], or a toroidal order [6]. So far, only the Bismuth ferrite $BiFeO_3$ has demonstrated room temperature multiferroic behaviour, and yet the complex antiferromagnetic ordering yields to a very small remnant magnetisation [7].

To face both these limitations, we decided to investigate completely different networks, namely the Tetragonal Tungsten Bronze (TTB) structural family. At the crystal chemistry level, this framework has two main advantages. First, the number of cationic sites is much higher than in perovskites (5 instead of 2) thus enabling extended substitutions opportunities to foster magnetic interactions. Next, TTB compounds are known to display incommensurate polar state [8,9], that may favour a coupling between magnetic and ferroelectric order. On investigating an extended family of TTB compounds, this paper is an experimental attempt towards this search for alternative multiferroics.

The "Tetragonal Tungsten Bronze" (TTB) structural type is related to the Potassium tungstate $K_{0.475}WO_3$, the structure of which was elucidated by Magneli [10]. The characteristic feature of the TTB crystal structure is the three types of open channels that develop within its octahedral framework (fig. 1). These features allowed solid state chemists to perform a wide range of substitutions, either in these channels or within the octahedral framework itself, which led to the discovery of $Ba_2NaNb_5O_{15}$ [11], one of the first and most significant representative of TTB niobates. The physical properties of $Ba_2NaNb_5O_{15}$ (Ferroelectricity, non-linear optics…) gave rise to extended investigations on TTB niobates, and to the discovery of many new ferroelectrics and more recently relaxors [12] compositions.

Besides this literature - and forty five years before the "revival of the magnetoelectric effect" [7] - Fang & Roth reported, in a short paper without experimental data [13], some of the title compositions as ferroelectric and ferrimagnetic materials, but did not report further investigations in later articles. A few years later, Krainik & al. [14] synthesized several rare-earth/iron substituted TTB niobates, and found some of them to be ferroelectrics. Schmidt, in his "magnetoelectric classification of materials" [15], referenced two of these TTB niobates. Ishihara & al. also reported on the TTB fluoride $K_3Fe_5F_{15}$ as a potential multiferroic [16].

On the basis of these works and of our own knowledge of TTB niobates, we decided to reinvestigate the $Ba_2LnFeNb_4O_{15}$ (Ln = La, Pr, Nd, Sm, Eu, Gd) system and started an in-depth study of rare-earth/iron substituted TTB niobates in order to confirm and understand their multiferroic properties.

In this paper, we report the synthesis and physical-chemical characterisation of several TTB niobates, with emphasis on their dielectric and magnetic properties. We observed that several compositions display both ferroelectric and magnetic hysteresis loops at room temperature. X-Ray microprobe experiments revealed that traces of barium ferrite $BaFe_{12}O_{19}$ are responsible for room temperature magnetic hysteresis. We also discuss the chemical, structural and physical properties of this family of TTB niobates in terms of crystal-chemical behaviour.

# Experiments

$Ba_2LnFeNb_4O_{15}$ (Ln = La, Pr, Nd, Sm, Eu, Gd) compounds have been obtained by conventional solid state route, from stoichiometric mixtures of $BaCO_3$, $Fe_2O_3$, $Nb_2O_5$ and $Ln_2O_3$, all reagents with 99.9% minimal purity grade. Ceramics were sintered at 1300°C and exhibited densities ranging from 90 to 96% of the nominal density of the products.

X-ray diffraction data were recorded on a Philips XPert pro diffractometer ($CuK_{\alpha 1}$, $\lambda$ = 1.54056 Å) with $10 < 2\theta < 130°$ and step = 0.008°. Rietveld refinements were performed using the program FULLPROF [17]. Experimental diffractograms were fitted using a Thompson-Cox-Hastings function (profile n°7), without any additional constraints.

Microstructural characterisations were performed on a JEOL 6560 SEM equipped with an EDS analyser.

X-Ray Microprobe mappings were obtained from a CAMECA SX-100 apparatus (15 kV, 10 µA).

Dielectric measurements were performed on a Wayne–Kerr 6425 component analyser under dry helium, using gold electrodes, at frequencies ranging from $10^2$ to $2.10^5$ Hz.

Ferroelectric hysteresis loops were performed using an analogical Sawyer Tower circuit with compensation of stray capacitances and resistances.

Magnetic hysteresis loops were obtained on a SQUID magnetometer (MPMS, Quantum Design Inc.).

## Results and Discussion

After standard solid state processing of powders and ceramics, X-Ray Diffraction (XRD) evidenced the successful stabilisation of the tetragonal tungsten bronze structure for all the compositions.

However the XRD diagrams of the Sm, Eu and Gd compounds displayed a few additional lines consistent with the presence of a Ferguson ite phase of formula $LnNbO_4$. Electron microscopy analysis confirmed the XRD results. We investigated various elaboration procedures in order to avoid this secondary phase formation. This allowed us to observe the limited thermal stability of TTB ceramics, which start decomposing at about 1360°C with formation of Hexagonal Tungsten Bronze (HTB) and Barium Ferrite, both identified during SEM experiments. However thanks to improved synthesis and sintering steps during the processing, the amount of Ferguson ite could be decreased in Sm, Eu and Gd samples. HTB and Barium Ferrite secondary phases remained undetectable from XRD and SEM analysis. In addition we point out that Ferguson ite phases have no particular dielectric and magnetic properties in the investigated temperature range.

Rietveld refinements were performed using an averaged model with a "pseudo-tetragonal" symmetry (space group Pba2, n°32). The refinements converged to satisfying agreement factors (Table 1, $R_B \approx 6$, $\chi^2 \approx 2$) and confirmed the stabilisation of the TTB crystal-structure. The cell parameters are reported in table 1, the b parameter is not reported because a and b refined values were undifferentiated. The Rietveld plot for the Sm representative, in which the $SmNbO_4$ secondary phase is taken into account, is displayed as an example in figure 2 (corresponding atomic coordinates are gathered in table 2). One can notice significant residues in the difference pattern, which could not be taken into account by any alternative structural model. There are also some discrepancies of the $B_{iso}$ atomic displacement factors for the octahedral sites ($Nb^{5+}$ and $Fe^{3+}$ ions). These features, and the undifferentiated values of a and b orthorhombic cell parameters, suggest that in these TTBs the crystal structure may be modulated. However the "pseudo-tetragonal" structures, even if they are averaged ones, provide a significant amount of reliable information that allows us to extract the trends governing the crystal-chemistry of these TTB multiferroics.

In these materials the octahedral framework is statistically occupied by $Fe^{3+}$ and $Nb^{5+}$ ions while the $Ba^{2+}$ ions occupy the pentagonal channels. The rare earths occupy the square channels. It should be noted that in the case of the gadolinium representative, the Rietveld refinement shows that Ba and Gd atoms are distributed other both pentagonal and square tunnels.

The crystal structures observed in the $Ba_2LnFeNb_4O_{15}$ TTBs are related to the introduction of the rare-earth cations in the square-shaped tunnels (see fig. 1) which normally define a 12-coordination. The $Ln^{3+}$ ions are reluctant to adopt such an environment, and this results in displacements of the $O^{2-}$ ions defining the surrounding octahedra. The cations occupying the octahedral sites are almost unaffected by this distorsion. In particular, although the 4c general position allows any kind of displacement, the corresponding ($Nb^{5+}$, $Fe3^{+}$) cationic sublattice remains almost perfectly tetragonal. The accommodation of the rare earth thus induces distortions in the octahedral framework. If correlated, these distortions may induce a modulated crystal structure. We note that, even in absence of Ln substitutions, the TTB network itself is likely to produce modulated structures, as in the case of $Ba_2NaNb_5O_{15}$ [8,9].

The influence of the $Ln^{3+}$ ions on the TTB framework can be illustrated by the evolution of the cell volume (tab. 1). The ionic radius of lanthanides ions is known to decrease with increasing atomic number. Thus the cell volume of the TTB matrix should decrease accordingly, when going from La to Gd. This trend is observed while going from La to Sm representatives, but the following representatives, Eu and Gd, points out a different crystal chemical behaviour. In particular the increase in the cell volume of the Gadolinium composition is probably related to the Ba/Gd statistical distribution over both pentagonal and square channels suggested by the corresponding Rietveld refinement. Taking into account the presence of fergusonite secondary phases in Sm, Eu and Gd TTbs, these observations suggest that the amount of smaller rare-earths that can be accommodated in the TTB framework is limited. This limitation is probably related to the distortion induced by the accommodation of the rare earth, depending on its ionic radius. These statements allow separating rare earth substituted TTBs into two groups, on crystal-chemical basis: La-Nd (total or almost total accommodation of the rare earths) and Sm-Gd (partial accommodation of the rare earths and detection of $LnNbO_4$ from XRD)).

From Scanning Electron Microscopy, the ceramics are highly densified (compactness > 95%) with micronic particles shaped like faceted rods at the surface. EDS analysis indicated cationic ratios in agreement with the expected 2:1:1:4 Ba:Ln:Fe:Nb ratio associated with the $Ba_2LnFeNb_4O_{15}$ composition, although Ln ratio was slightly lower in Sm, Eu and Gd samples. Traces of fergusonite phase were detected in the Pr and Nd samples, suggesting that in both these materials the accommodation of the rare earth is not complete. Thus SEM results confirm that the tetragonal tungsten bronze structure has been successfully stabilized after the introduction of paramagnetic cations and indicate that the microstructure of the ceramics is optimal.

Dielectric measurements on the La sample did not reveal any ferroelectric transition down to 82K, although the evolution of the dielectric permittivity suggests a possible transition below 80 K, *i.e.* beyond our experimental conditions.

The Nd, Sm and Eu representatives are ferroelectric below $T_C$ = 323 , 405 and 440 K respectively, as illustrated by figure 3a. This is confirmed by ferroelectric hysteresis loops which were obtained at room temperature, i.e. below the transition temperature for the three samples (figure 4).

On the other hand, the Pr and Gd samples are relaxors-ferroelectrics below 170K and 300K respectively (Fig. 3b). This means that both the real and imaginary parts of the dielectric permittivity display a broad dielectric maximum whose temperature shifts on sweeping the frequency. Because of this relaxor-ferroelectric behaviour, no ferroelectric hysteresis loop could be recorded in these latter materials.

The details of the dielectric properties of all samples are gathered in table 2. One can note that the dielectric transition temperature and the maximum permittivity follow similar trends, although the Eu TTB displays a quite high permittivity. However the dielectric permittivities in Tetragonal Tungsten Bronze ferroelectrics can be affected by order/disorder phenomenon in the different kind of channels of this framework, and we suspect that this is the case for Eu TTB. The polarisations extracted from ferroelectric hysteresis loops for the Nd, Sm and Eu TTBs are also rather low, as compared to the usual polarisations observed in TTB ferroelectrics (10-20 $\mu C.cm^{-2}$). This could be related to a modulated structure (suggested by Rietveld refinements) that induces misalignments of the local dipoles and thus reduces the overall polarisation.

The dielectric studies show that a ferroelectric behaviour can be obtained at room temperature and above in some of these magnetically substituted TTBs. Rietveld studies indicate that the electric dipoles in the Nd, Sm and Eu samples originate from an off-center position of the $Nb^{5+}$ ions within the octahedral sites. All these sites display one short and one long axial bond length, in agreement with a statistical distribution of niobium and iron over the whole octahedral framework. Off-center positions of the $Nb^{5+}$ ions are also observed in the relaxor $Ba_2GdFeNb_4O_{15}$, but display inhomogeneous distortions in the different octahedral sites.

The evolution of the ferroelectric properties in the Nd, Sm and Eu compounds can be rationalised from a crystal-chemical point of view, which could be resumed as follows:
- The smaller the rare earth, the greater the distortion of the TTB framework
- The greater the distortion of the TTB framework, the greater the permittivity
- The greater the distortion of the TTB framework, the higher the Curie temperature

These considerations indicate that the accommodation of the rare earth in the TTB framework is the critical parameter impacting the dielectric properties in this family of materials.

After this description of ferroelectric properties we will now focus on the magnetic properties observed in these samples at room temperature.

Magnetic hysteresis loops at 300 K have been obtained from all the samples except the La one (fig 5). The 20 kOersteds magnetisations are comparable, while the coercive fields are about 500-700 Oersteds for Pr, Eu and Gd samples and about 2000 Oe for Nd and Sm samples. It should be noted that all the samples, except the La one, are strongly attracted to a NdFeB magnet at room temperature.

The details of the dielectric and magnetic properties of the studied samples are gathered in table 3. It is worth mentioning that the room temperature magnetizations of the TTB samples evolve in a similar way as the dielectric transition temperature. As was observed for the dielectric properties, the variation of the magnetic properties with respect to the rare earth shows that the $Ln^{3+}$ ions play a critical role concerning the magnetic behaviour of our TTBs samples.

The magnetic study suggests a successful magnetic substitution, as all the samples, except the La one, exhibit hysteresis loops at room temperature. The existence of a spontaneous magnetisation at room temperature in these samples, however, is surprising considering that:
- The $Ln^{3+}$ ions are not likely to be magnetically ordered at room temperature
- The $Fe^{3+}$ ions are diluted and disordered in the $[NbO_6]$ octahedral framework

This latter statement in particular, is not in favour of a long range magnetic order. Thus we will now discuss the possible links between the crystal-chemistry and the magnetic and ferroelectric properties of these niobates, which may account for the origin of the magnetic properties.

The trends in magnetic properties, as well as the evolution of dielectric properties, can be correlated to the evolution of cell volume. As shown by figure 6, the molar magnetisation of the TTB samples, measured at room temperature under a magnetic field of 20 kOe, and the dielectric ordering temperature evolve inversely proportional with respect to the cell volume. This correlation is reproducible and is observed for several sets of samples. Thus the correlation between the magnetic behaviour of TTBs and their structural and dielectric properties may support an intrinsic origin for the multiferroic character of the $Ba_2LnFeNb_4O_{15}$ TTBs (Ln = Pr, Nd, Sm, Eu, Gd).

However X-ray microprobe analysis, initially performed in order to determine more precisely the rare-earth composition in fergusonite-containing samples, revealed traces of a Fe-rich secondary phase. This parasitic phase is not detected within the La TTB (fig. 7b and 7a respectively), which also do not exhibit magnetic hysteresis loops at room temperature. Moreover the Fe-rich phase contains a small amount of barium, suggesting a barium ferrite $BaFe_{12}O_{19}$ composition, *i.e* a widely used permanent magnet. Thus the magnetic hysteresis loops observed at room temperature in most of the $Ba_2LnFeNb_4O_{15}$ samples originate from traces of barium ferrite.

The presence of barium ferrite, and its amount, can be related to the accommodation of the rare earth in the TTB matrix. In the La sample, the $La^{3+}$ ion is totally accommodated within the TTB framework, and no fergusonite phase $LaNbO_4$ appear. To maintain the charge balance of the TTB phase, $Fe^{3+}$ ions must be totally incorporated in the TTB framework of the La sample. In the other samples (Ln = Pr, Nd, Sm, Eu, Gd) the rare earths are partly accommodated, and the excess of lanthanide oxide reacts with niobium oxide to form $LnNbO_4$ fergusonite. Thus, to maintain electrical neutrality of the TTB framework in these cases, iron must be only partly accommodated in the TTB framework. The excess of iron oxide reacts with a small proportion of barium oxide to form barium ferrite related phases. It appears that the nature of the rare earth, and particularly its size, *i.e.* its ability to be accommodated within the TTB framework, controls the appearance of the magnetic barium ferrite phase in our samples.

This latter correlation indicates that the rare earth introduced in the TTB framework, *in fine*, has a direct influence on its chemical stability, its crystal-chemistry and its physical properties. Beyond the correlation shown in fig. 6, one has to take into account the additional and highly critical parameter of the chemical stability of these TTB compositions.

Besides this correlation between chemical stability, crystal-chemistry and physical properties, it is worth mentioning that all the TTB samples (except the La one), display both spontaneous polarisation and magnetisation and can be considered as composites multiferroics. The particular interest of these multiferroic composites is that the magnetic phase is generated in situ, during the elaboration of the sample, which should favour the quality of the interfaces between the ferroelectric and magnetic part of the composite. Adequate chemical substitutions can also modify the amount and properties of this magnetic phase. Moreover, four of these composites materials display both ferroelectric and ferromagnetic hysteresis loops at room temperature. The particular case of the Pr and Gd samples should be noted, for they are, below 170 and 300 K respectively, two examples of composite relaxor multiferroics. Last but not least, this work shows that it is possible to introduce magnetic cations in a ferroelectric framework without loosing the ferroelectric properties.

# Conclusion

The experiments reported in this paper demonstrate that five over the six studied Tetragonal Tungsten Bronze samples are composite multiferroics below their ferroelectric/relaxor Curie temperature. The multiferroic properties of these materials are directly influenced by the nature of the rare earth accommodated in the TTB framework. Thus the Nd, Sm, Eu and Gd representatives are multiferroic at room temperature and above. Moreover the Gadolinium and Praseodymium representative are relaxor multiferroics. The $Ba_2LnFeNb_4O_{15}$ niobates (Ln = Nd, Sm, Eu, Gd) constitute a family of room-temperature composite multiferroics. Both the composite and single-phase approaches are still under investigation.

We would like to point out again our initial approach, which was to introduce new crystalline frameworks in the field of multiferroic materials. We hope that this work will contribute to the widening of the crystal-chemical aspects in multiferroic materials, for such an approach may provide a substantial support to the design of single-phased multiferroics.

# Acknowledgements

The authors want to thanks L. Raison for the X-Ray microprobe experiments, and E. Lebraud and S. Pechev for the collection of X-Ray diffraction data.

This work is supported by the European Network of Excellence "FAME" ([www.famenoe.net](www.famenoe.net)) and the STREP MaCoMuFi ([www.macomufi.eu](www.macomufi.eu)).

**Figures**

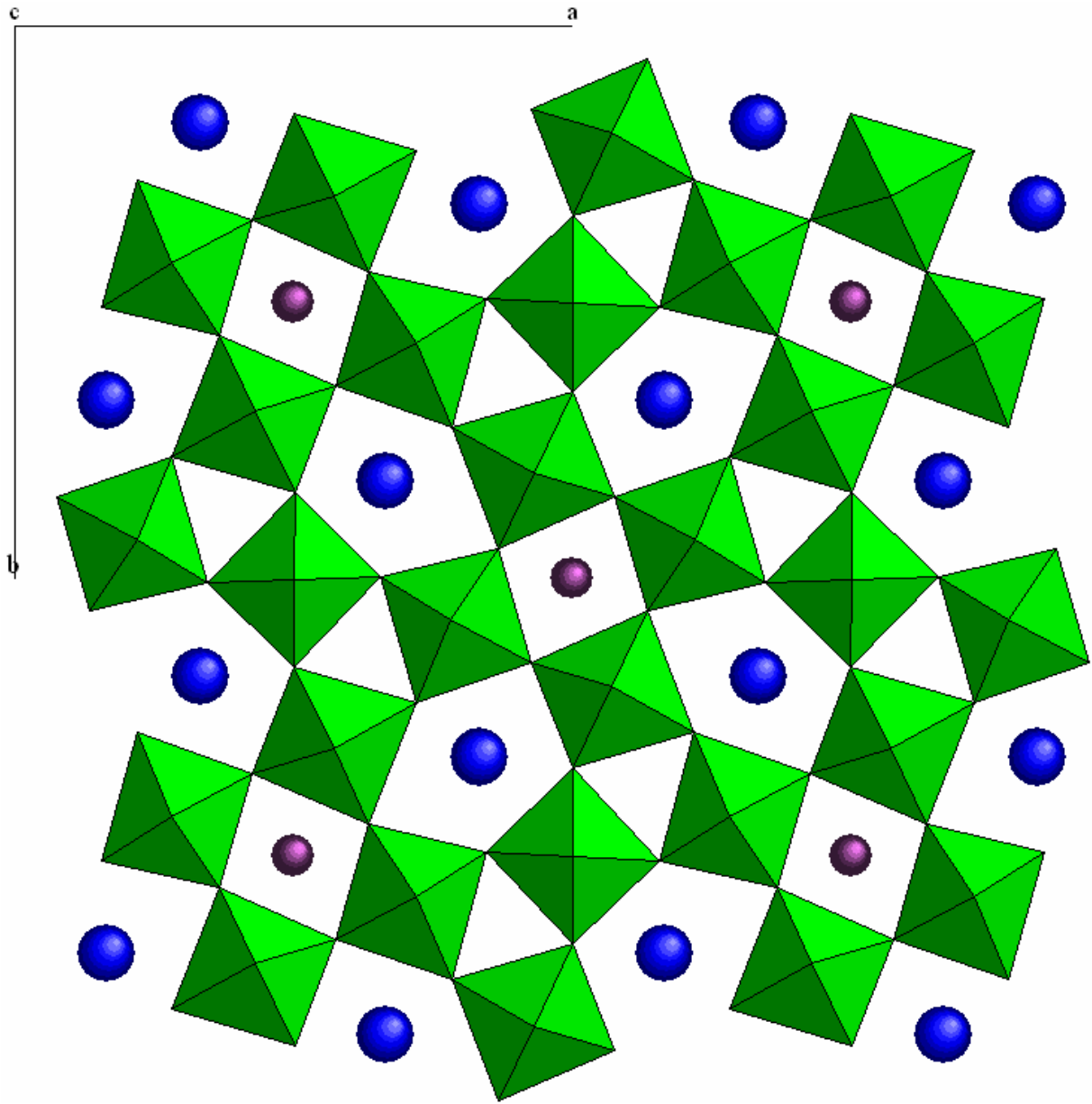

**Figure 1**

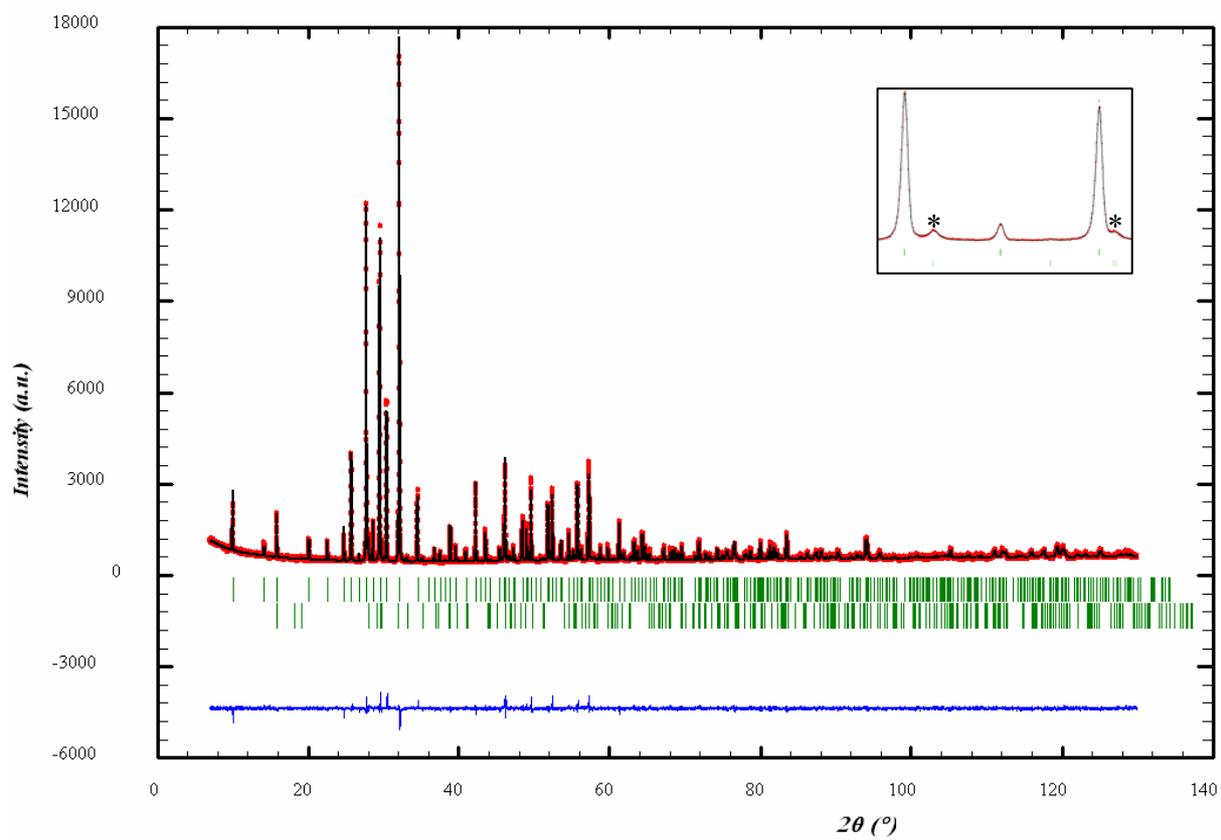

**Figure 2**

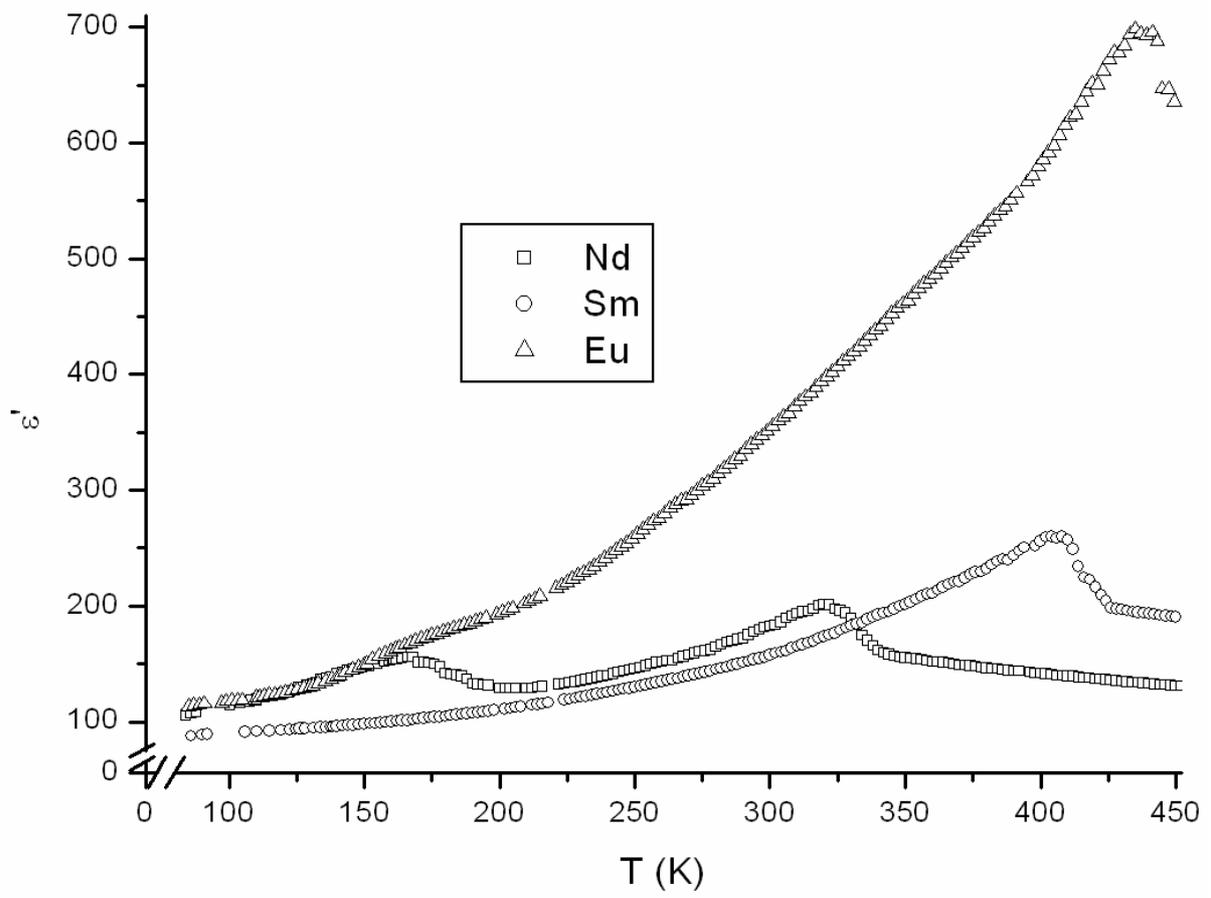

**Figure 3a**

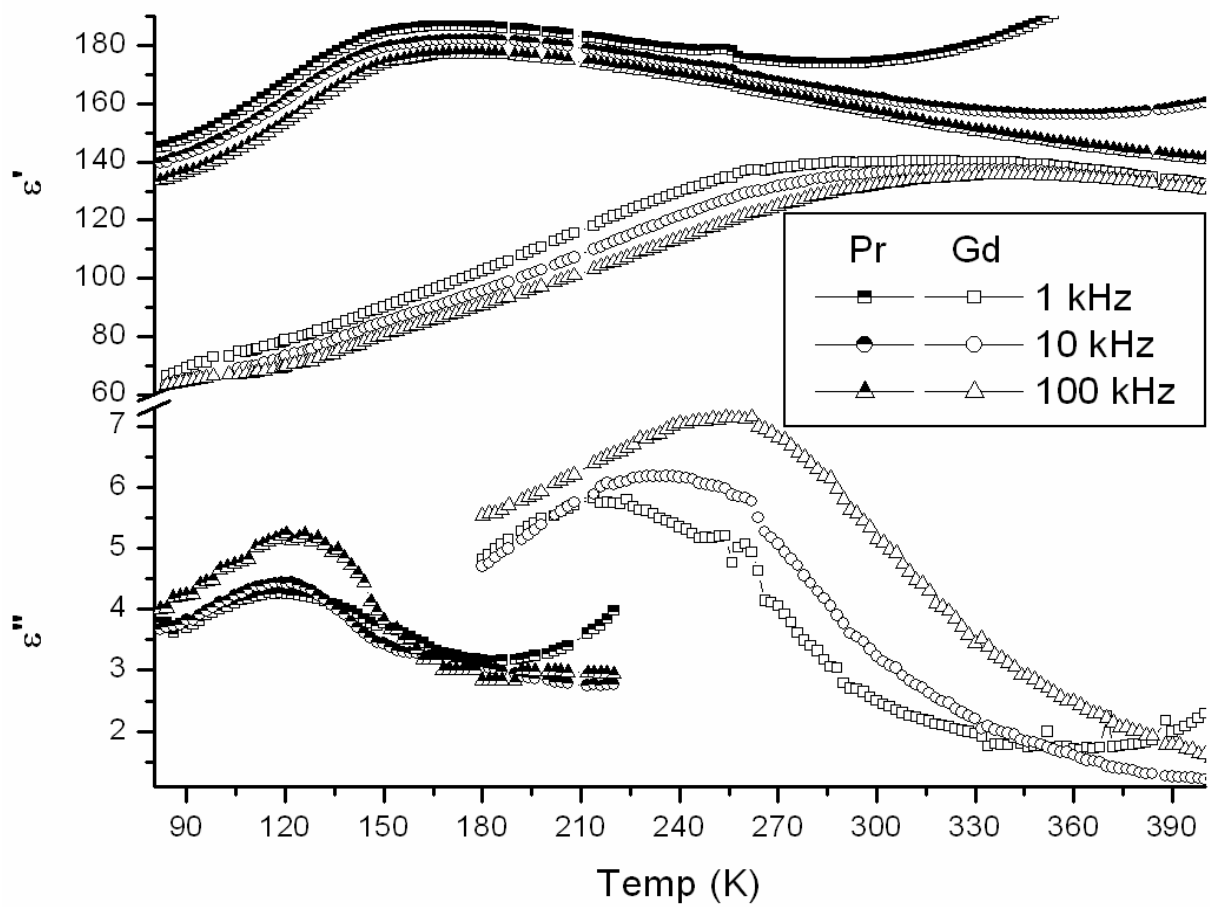

**Figure 3b**

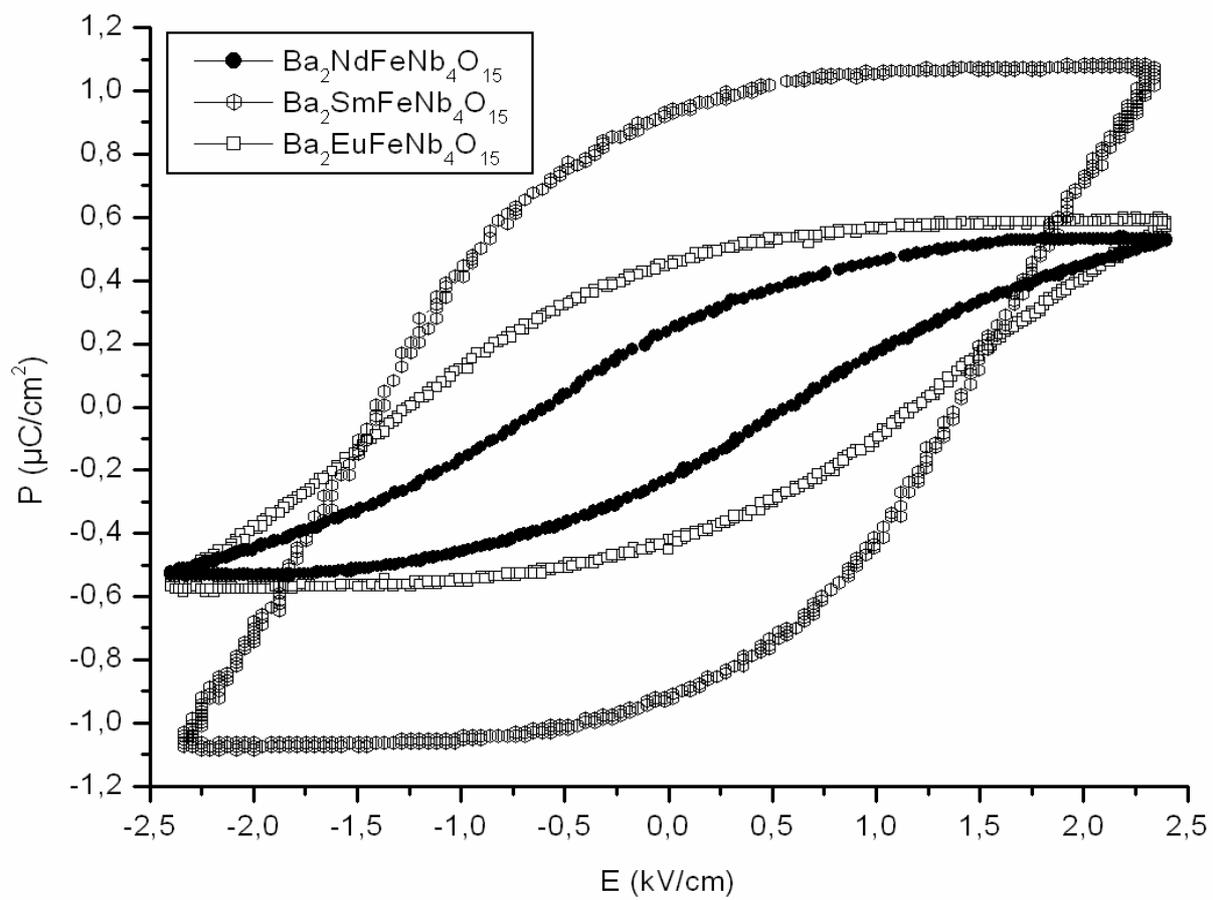

**Figure 4**

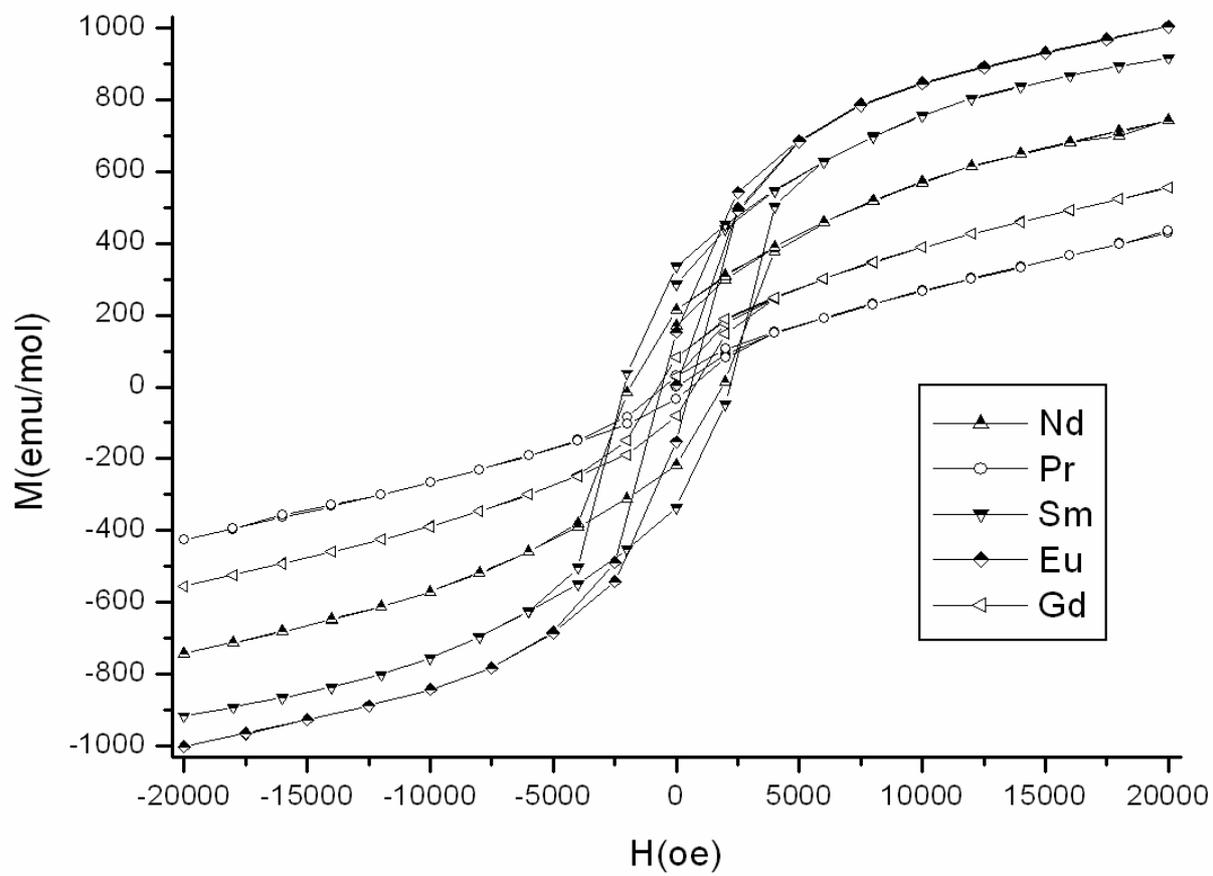

**Figure 5**

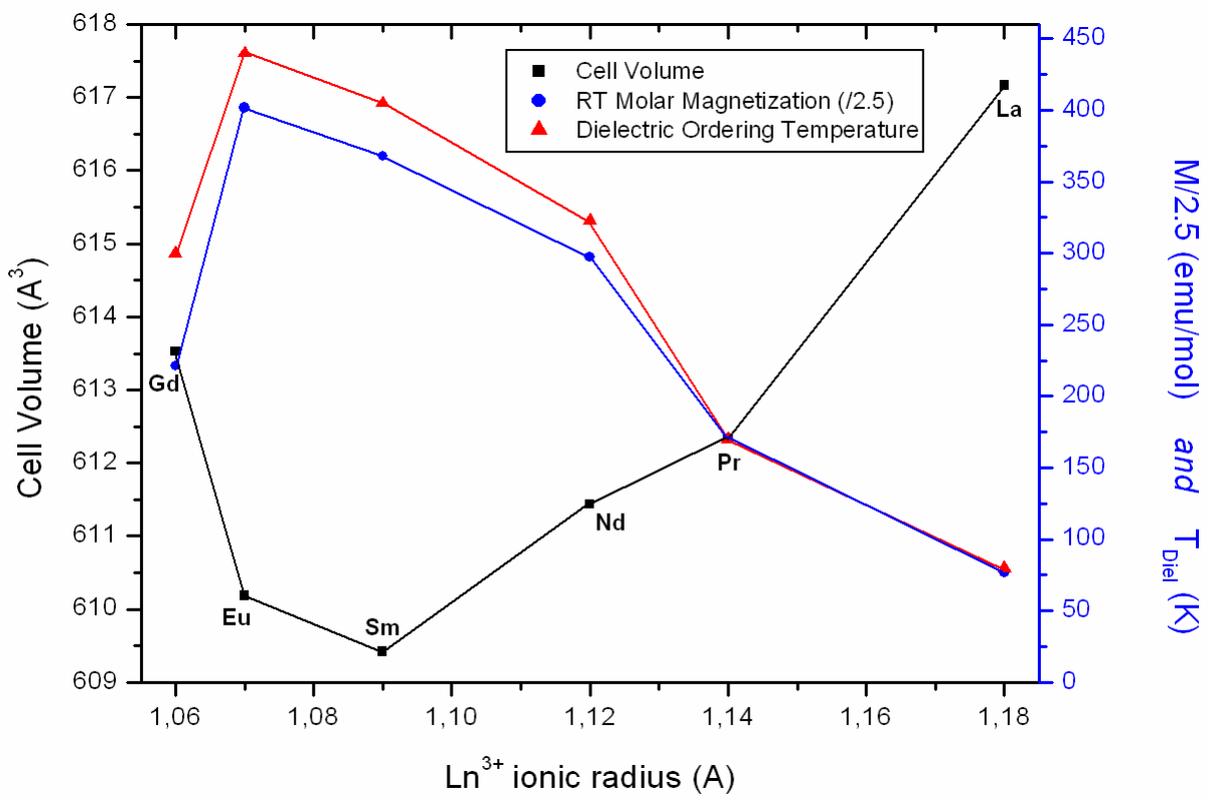

**Figure 6**

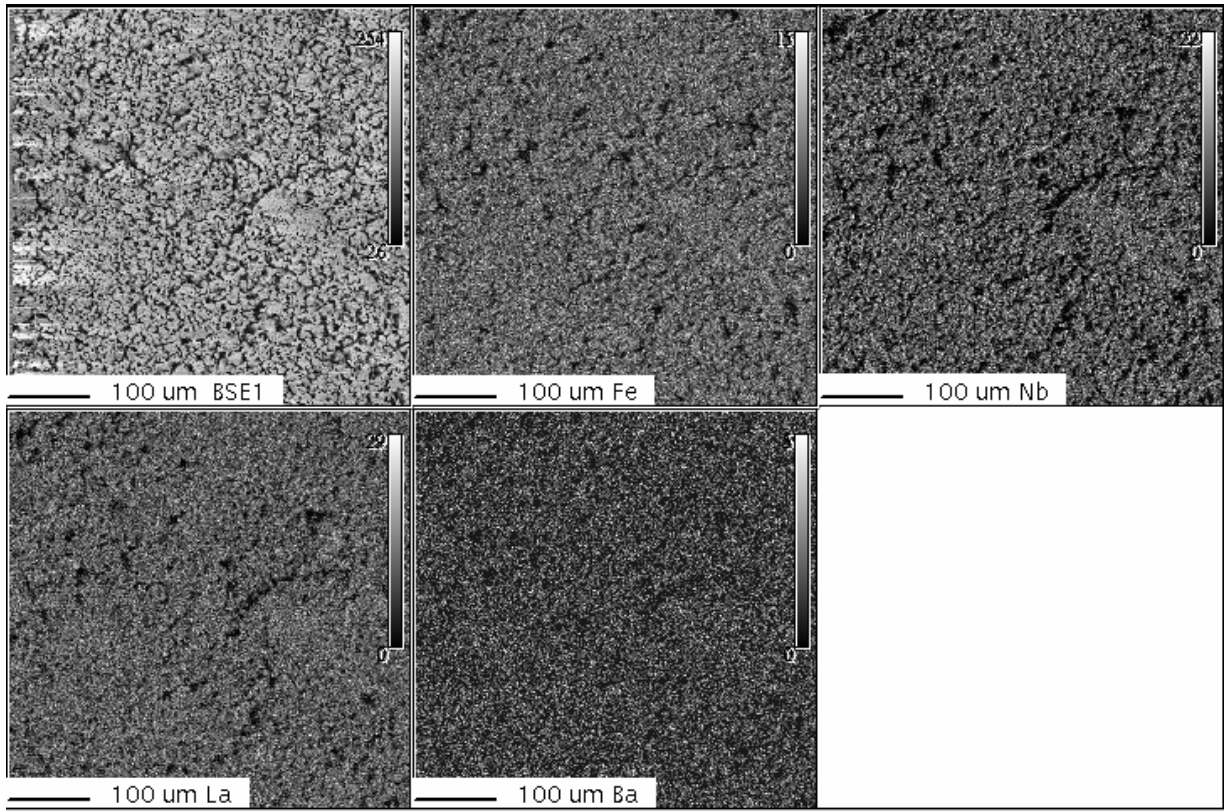

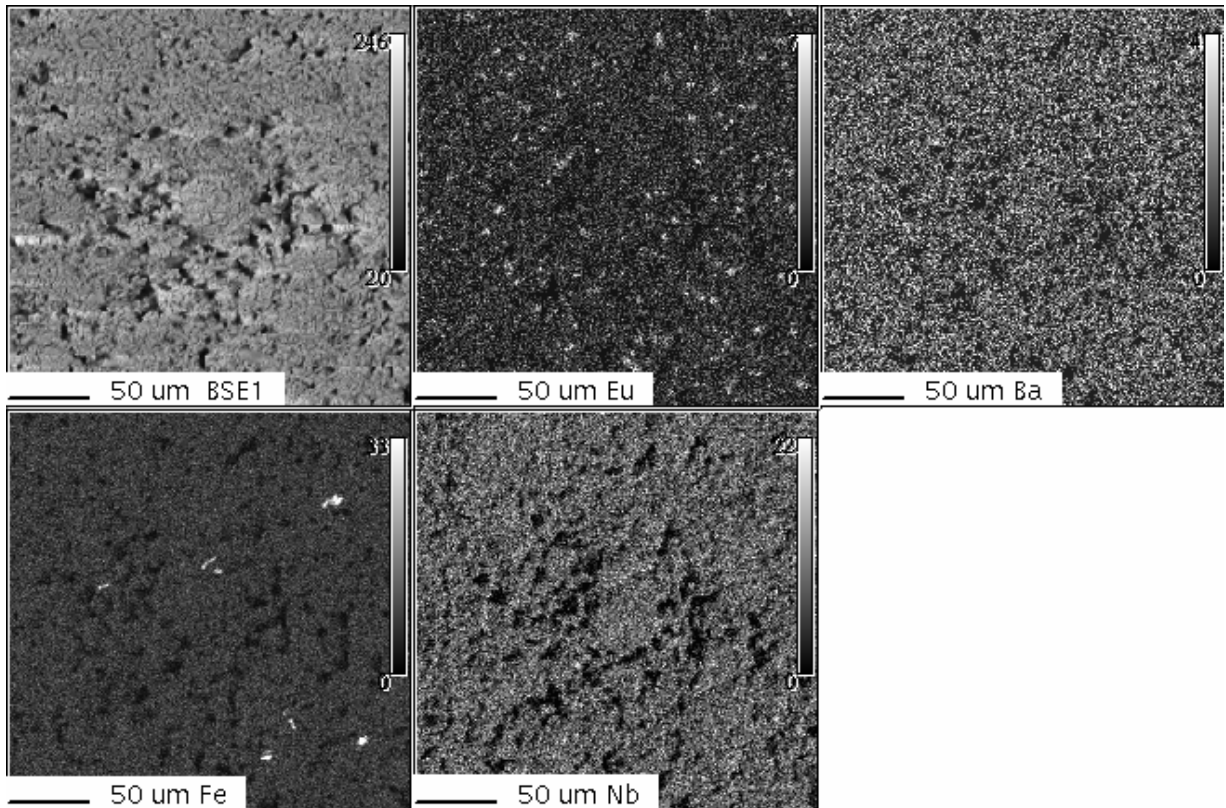

**Figure 7**

# Figures captions

Figure 1: Crystal structure of $Ba_2LnFeNb_4O_{15}$ viewed along the c axis, showing the distorsion of the square channel.

Figure 2: Rietveld Plot for $Ba2SmFeNb_4O_{15}$ (dots: experimental, continuous line: calculated), insert shows $SmNbO_4$ contributions (*). Upper vertical bars indicate TTB Bragg positions (lower bars: $SmNbO_4$).

Figure 3: a) Dielectric properties of $Ba_2LnFeNb_4O_{15}$ TTBs, Ln = Nd, Sm, Eu at 10kHz b) Ln = Pr, Gd, above vertical scale break: real part of the dielectric constant, below scale break : imaginary part of the dielectric constant (for clarity only the relevant sections of the imaginary part are displayed).

Figure 4: Ferroelectric hysteresis loops at room temperature for the Nd, Sm and Eu TTBs

Figure 5: Magnetic hysteresis loops at room temperature for the $Ba_2LnFeNb_4O_{15}$ (except La) TTBs

Figure 6: Evolution of unit cell volume, molar magnetization (RT, H = 20kOe) (divided by 2.5 for scale purpose) and dielectric ordering temperature of $Ba_2LnFeNb_4O_{15}$ TTBs versus ionic radius of the rare earth.

Figure 7: X-Ray microprobe mappings of a $Ba_2LaFeNb_4O_{15}$ (a) and $Ba_2EuFeNb_4O_{15}$ (b) TTB samples (BSE : backscattered electron image of the mapped zone). White dots in the Eu and Fe cartography of $Ba_2EuFeNb_4O_{15}$ correspond to $EuNbO_4$ and barium ferrite, respectively.

| Ln | a (Å) | c (Å) | V (Å$^3$) | R$_B$ | R$_p$ | R$_{wp}$ | R$_{exp}$ | $\chi^2$ |
|---|---|---|---|---|---|---|---|---|
| La | 12.514(1) | 3.941(1) | 617.2 | 5.74 | 15.1 | 13.7 | 10.35 | 1.77 |
| Pr | 12.484(1) | 3.928(1) | 612.3 | 5.87 | 15.4 | 16.6 | 10.07 | 2.73 |
| Nd | 12.478(1) | 3.925(1) | 611.0 | 6.45 | 21.9 | 19.4 | 12.65 | 2.39 |
| Sm | 12.458(1) | 3.928(1) | 609.8 | 5.83 | 17.0 | 13.6 | 11.76 | 1.35 |
| Eu | 12.460(1) | 3.930(1) | 610.2 | 7.06 | 19.8 | 17.0 | 11.55 | 2.17 |
| Gd | 12.485(1) | 3.936(1) | 613.5 | 6.24 | 21.7 | 17.4 | 13.27 | 1.74 |

**Table 1**

| Atom | Wickoff | occupancy | x | y | z | $B_{iso}$ (Å²) |
|---|---|---|---|---|---|---|
| Ba | 4c | 1.000 | 0.1656(2) | 0.6768(2) | 0.3510(6) | 0.77(3) |
| Sm | 2a | 1.000 | 0 | 0 | 0.3480(7) | 0.81(3) |
| Nb1 | 2b | 0.800 | 0 | 1/2 | 0.8713(5) | 0.45(5) |
| Fe1 | 2b | 0.200 | 0 | 1/2 | 0.8713(5) | 0.45(5) |
| Nb2 | 4c | 0.800 | 0.2847(4) | 0.4267(3) | 0.8555(5) | 0.02(2) |
| Fe2 | 4c | 0.200 | 0.2847(4) | 0.4267(3) | 0.8555(5) | 0.02(2) |
| Nb3 | 4c | 0.800 | 0.5741(3) | 0.2878(4) | 0.8510(6) | 0.02(2) |
| Fe3 | 4c | 0.200 | 0.5741(3) | 0.2878(4) | 0.8510(6) | 0.02(2) |
| O1 | 4c | 1.000 | 0.163(2) | 0.499(2) | 0.753(3) | 0.70(13) |
| O2 | 4c | 1.000 | 0.490(2) | 0.157(2) | 0.850(2) | 0.70(13) |
| O3 | 4c | 1.000 | 0.291(1) | 0.775(1) | 0.813(2) | 0.70(13) |
| O4 | 4c | 1.000 | 0.351(1) | 0.575(2) | 0.733(3) | 0.70(13) |
| O5 | 4c | 1.000 | 0.436(1) | 0.378(1) | 0.885(1) | 0.70(13) |
| O6 | 4c | 1.000 | 0.211(1) | 0.950(1) | 0.358(2) | 0.70(13) |
| O7 | 4c | 1.000 | 0.092(1) | 0.190(1) | 0.403(2) | 0.70(13) |
| O8 | 2b | 1.000 | 0 | 1/2 | 0.346(3) | 0.70(13) |

**Table 2**

|  | La | Pr | Nd | Sm | Eu | Gd |
|---|---|---|---|---|---|---|
| **Dielectric prop.** | - | Rel. | Ferroel. | Ferroel. | Ferroel. | Rel. |
| $T_C$ or $T_m$ (K) | - | 170 | 323 | 405 | 440 | 300 |
| $\varepsilon'_{max}$ (1 kHz) | 178 | 186 | 206 | 266 | 731 | 137 |
| $P_S$ (µC/cm$^2$) | - | - | 0.53 | 1.08 | 0.59 | - |
| **Magnetic prop. (RT)** | - | Hyster. | Hyster. | Hyster. | Hyster. | Hyster. |
| $H_c$ (Oe) | - | 540 | 1860 | 2150 | 580 | 700 |
| $M_{(2T)}$ (e.m.u. / mol) | 192.8 | 430.0 | 741.9 | 916.2 | 1004.5 | 554.3 |

**Table 3**

**Tables captions**

Table 1: Cell parameters (b ≈ a) and volume, agreement factors of the Rietveld refinement.

Table 2: Atomic coordinates and isotropic atomic displacement parameters for $Ba_2SmFeNb_4O_{15}$

Table 3: Dielectric and magnetic properties of $Ba_2LnFeNb_4O_{15}$ samples ($\varepsilon'_{max}$ taken at $T_C$ or $T_m$ except for La TTB (80K))